\documentstyle[aps,twocolumn]{revtex}

\def\Ef{E_{\mbox{\tiny F}}} \def\vf{v_{\mbox{\tiny F}}}

\begin{document}
 
{\large \bf Reply to Comment on ``Scaling of the quasiparticle
  spectrum for $d$-wave superconductors''}

Volovik and Kopnin\cite{Volcom} have raised two separate issues in the
preceding comment -- (1) that the linearization of the quasiparticle
spectrum around the gap node limits the maximum temperature for which
the discussion in Ref. \onlinecite{SimonLee} is valid, and (2) that
the use of the geometric average light cone velocity in Ref.
\onlinecite{SimonLee} gives an incorrect estimate of the crossover
scale.  Both of these objections are valid and will be discussed
below.  We note that, with the exception of this incorrectly predicted
crossover scale, the results of our paper remain unchanged.

We first address the issue of linearization of the quasiparticle
spectrum.  As mentioned in \onlinecite{SimonLee} the validity of the
linearized dispersion is expected to be restricted to a temperature
range $T \ll \Delta^2/\Ef$ (with $\Delta \sim T_c$ is the maximum gap)
where excitation energies are small enough such that the quadratic
part of the Hamiltonian (the part representing curvature of the Fermi
surface) is much smaller than the leading linearized piece.  On the
other hand, the form of the quasiparticle spectrum described by
Volovik and Kopnin (Eq. 1 of Ref.  \onlinecite{Volcom}) accounts for
the curvature of the Fermi surface and therefore can be used to
describing excitations at energy scales up to the order of
$\Delta^2/\Ef$ for quasiclassical calculations for which $\vec p$ is
considered to be a good quantum number\cite{Vol1,Vol2}.

It should be noted however that as mentioned in \onlinecite{SimonLee},
in practice, the Fermi surface in the high $T_c$ compounds can be
quite flat at the gap nodes such that the Fermi surface curvature is
smaller than one would expect for a model circular Fermi surface, and
thus the range of validity of the linearization used in
\onlinecite{SimonLee} may be somewhat larger than otherwise expected.
It should also be noted that when curvature of the Fermi surface is
important (such as for the calculation of the thermal Hall
coefficient), it can be effectively treated using perturbation
theory\cite{SimonLee}.

We now turn to the separate issue of the crossover scale.  Volovik and
Kopnin point out two crossover scales $T_1 \sim T_c \sqrt{H/H_{c2}}$
and $T_2 \sim (T^2_c/\Ef) \sqrt{H/H_{c2}}$.  In the absence of
magnetic field, the velocity in the direction perpendicular to the
Fermi surface is the Fermi velocity $\vf$ whereas the velocity
tangential to the Fermi surface is roughly $\sim \vf T_c/\Ef$.  In a
magnetic field, the periodicity of the vortex lattice is given by the
magnetic length $l_0$, thus the Brillouin zone edge is at momentum
$k_{max} \sim 1/l_0$.  Neglecting the vector potential and assuming
that momentum remains a good quantum number, we find that the energies
of the states at the zone edge in the two different directions
correspond to the two crossover energy scales discussed above.  In
Ref.  \onlinecite{SimonLee} it was incorrectly assumed that in a
magnetic field, the semiclassical states precess thus obtaining a
single geometrically averaged velocity between the two different
directions.  However, as correctly treated in \onlinecite{Vol2}, the
Eilenberger semiclassical approach leaves $\vec p$ a good quantum
number even in a magnetic field so that the momentum states do not
actually precess.  This result can also be seen from the dynamics of
the linearized Hamiltonian in \onlinecite{SimonLee}.

In fact, however, the situation is somewhat more complicated than the
above paragraph would lead us to believe.  When we add a magnetic
field\cite{SimonLee}, one component of the vector potential (times
$\vf$) acts as a periodic scalar potential for delocalized
quasiparticles, resulting in a gap at the zone edge of size $T_1$ in
{\it both} direction in the Brillioun zone.  Whether a gap is actually
observed in the density of states depends on the details of the band
structure.  Nonetheless, it is clear that this should be an important
crossover scale being that this is also the typical energy of the
periodic potential.

The prediction of the additional crossover scale at energy $T_2$ is
due to the somewhat different physics of the bound vortex core states.
Volovik and Kopnin\cite{Volcom,Vol1,Vol2} have calculated that the
spacing of the core states is approximately $T_2$ at low energy.  In
this low energy range, however, the major contribution to the density
of states is from the delocalized quasiparticles\cite{Vol1}, so we
would probably only see this discretization clearly if a gap occurs in
the spectrum of the extended states.  Thus, an important direction for
future research will be to attempt an exact quantum mechanical
treatment of the spectrum at these low energies.

This work was supported by NSF Grant No. DMR-95-23361

\vspace*{10pt}

\noindent Steven H. Simon and Patrick A. Lee 
\\ \hspace*{5pt}
\begin{minipage}[t]{3in} Department of Physics \\ Massachusetts
  Institute of Technology \\ Cambridge, Massachusetts 02139
\end{minipage}

\vspace*{10pt}

\noindent \today 

\end{document}